# Plasmonic Hot–Carrier Extraction: Mechanisms of Electron Emission

Charlene Ng,[*,†] Peng Zeng,[‡] Julian A. Lloyd,[¶] Debadi Chakraborty,[§] Ann Roberts,[◊] Trevor A. Smith,[‡,⊥] Udo Bach,[¶,⊥] John E. Sader,[§,⊥] Timothy J. Davis,[◊] Daniel E. Gómez [*,#,†,⊥]

†CSIRO, Manufacturing, Private Bag 33, Clayton, VIC, 3168, Australia
‡School of Chemistry, The University of Melbourne, VIC 3010, Australia
¶Department of Materials Science and Engineering, Monash University, Clayton, VIC 3800, Australia
§School of Mathematics and Statistics, The University of Melbourne, VIC 3010, Australia ◊School of Physics, The University of Melbourne, VIC 3010, Australia
⊥ARC Centre of Excellence in Exciton Science
#RMIT University, Melbourne, VIC, 3000, Australia

When plasmonic nanoparticles are coupled with semiconductors, highly energetic hot carriers can be extracted from the metal–semiconductor interface for various applications in light energy conversion. Hot charge–carrier extraction upon plasmon decay using such an interface has been argued to occur after the formation of an intermediate electron population with a uniform momentum distribution. The efficiency of the charge separation process is thus discussed to be limited by this spatial homogeneity in certain plasmon-induced applications. Here we demonstrate using visible pump, near–infrared probe transient absorption spectroscopy that increases in the contact area between metal and semiconductor leads to an increase in the quantum yield for hot electron injection that is inconsistent with the homogeneous energy–momentum distribution of hot–electrons. Instead, further analysis of the experimental data suggests that the highly energetic electrons are emitted across the interface via a surface charge emission mechanism that occurs via a plasmon excitation.

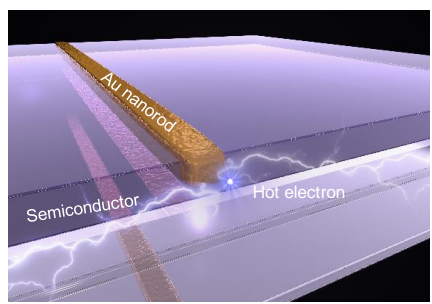

Nanoscale heterogeneous structures consisting of metal-semiconductor junctions have great potential in applications related to photonic energy conversion. When plasmonic metal nanostructures are in direct contact with a semiconductor, a Schottky barrier can be formed at the metal–semiconductor interface[2] [Figure 1(A)]. Hot charge carriers that result from non–radiative plasmon decay can be emitted across this interface, leading to charge–separated states that are useful for various applications, including photodetectors[3], photovoltaics[4] and photochemistry.[5] The reported collection efficiency of these electrons is low, typically < 1%.[3,4,6-9] However, Schottky photodetectors are valuable due to their fast response and low photon-energy detection capability.[1] Furthermore, the operational lifetime of photocatalytic Schottky devices surpasses that of more efficient semiconductors[7].

It has been postulated that non–radiative plasmon relaxation results in a transient population of energetic electrons with a uniform spatial distribution of momenta.[10,11] This spatial uniformity limits the efficiency of charge–separation in metal–semiconductor junctions.[12] Typically, the metal–semiconductor interface consists of a single, planar junction[3,4] [Figure 1B] where only those electrons that simultaneously move toward the single interface and have sufficient kinetic energies to overcome potential energy barriers can be emitted (shaded area in the momentum–space sphere shown in Figure 1(B)). Simplified models predict that, for moderate electron energies, only a small fraction of the total hot–electron population resulting from plasmon decay meet these criteria, leading to poor quantum yields for hot–electron extraction.[13]

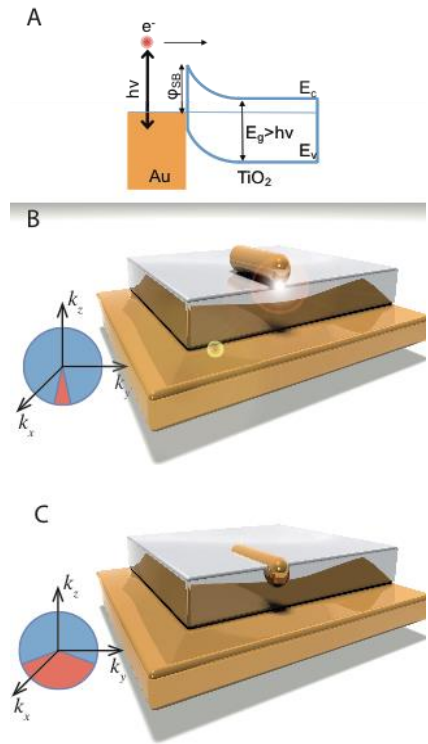

**Figure 1. *Partially-embedded Au nanorods.*** *(A) Band diagram of a metal–semiconductor contact for the extraction of plasmonic hot-charge carriers. $E_g$, $E_c$ and $E_v$ are the bandgap, conduction band and valence band of the semiconductor material respectively. $\varphi_{SB}$ is the Schottky barrier[1]. Absorption of a photon of energy hv ($< E_g$) and non-radiative relaxation could lead to electron injection from the metal to the semiconductor. (B) Metal-on-semiconductor structure: On momentum space, there is a narrow escape cone for electron injection. (C) When the metal nanostructure is partially embedded, the number of interfaces available for electron injection increases.*

Other limiting factors for hot-electron extraction stem from properties intrinsic to metals, including ultra-fast carrier-carrier relaxation pathways[14] and broad energy distributions of hot-carriers, which stem from an almost continuous density of electronic energy states that exist below the Fermi level of the metal.[10] These limitations can be overcome via a direct charge-transfer mechanism,[15] which can take place when the metallic nanostructure and the semiconductor material have a strong electronic interaction[16], a situation that has led to quantum yields for charge separation of >

20%.[15] Another approach to increase the charge–separation quantum yield is to increase the contact area between metal and semiconductor [Figure 1(C)].[9] This in principle allows for a larger fraction of the excited hot electron population to be extracted into the semiconductor material (shaded area in the momentum–space sphere shown in Figure 1(C)). Whilst this concept was demonstrated in the past by[9], their approach is of limited scalability due to the use of complex top–down nanofabrication techniques. Furthermore, top–down fabrication approaches set a lower–limit to the minimum size of the metal nanostructures (> 20 nm), which in turn set the optical response of the nanostructures to the NIR region of the spectrum. This implies a low electrochemical potential of the the charge–separated state limiting the range of applications in hot–electron photo–redox catalysis.[17] An important point to note is that full coverage of the plasmonic nanostructures by the semiconductor could hinder hot–carrier extraction.

Here we demonstrate that the increase in the metal–semiconductor contact area result in augmented yields for electron injection that are inconsistent with the uniform distribution of electron momenta. We achieve the increased metal–semiconductor contact area by means of a simple, large scale and low cost fabrication process, using nanocrystaline Au rods as the building blocks. The electron injection yields are estimated from ultra–fast pump–probe spectroscopy measurements, which demonstrate a 2.7× increase in the quantum yield for electron injection with contact area. This increase is consistent with a surface charge emission mechanism that occurs via a plasmon excitation.

In Figure 2 we show a schematic of the "lift–off" process we have developed to create partially embedded metal–semiconductor structures over macroscopic areas. Firstly

(step A, Figure 2), a thin layer of methyl methacrylate (MMA, an acetone–soluble material) is spin–coated onto a cleaned Si wafer and Au nanorods are self–assembled (by means of electrostatic interactions) on the surface of MMA by immersing the MMA/Si wafer in an aqueous dispersion containing Au nanorod colloid (more details in the methods section). After the Au nanorods are self–assembled onto the MMA/Si wafer, a layer of titanium dioxide ($TiO_2$, 50 nm) is deposited onto the structure via a standard electron beam evaporation process (step B, Figure 2). The thickness of this layer must exceed the diameter of the Au nanorods (estimated to be ~ 20 nm, supporting information figure S1) in order to ensure almost complete coverage of the nanostructures. At this point, more layers could be added, including a mirror, which would result in the creation of metal–semiconductor–metal structures that can exhibit increased light absorption[18,20] (supporting information section S2). An epoxy adhesion layer is then used to bond the complete structure to a glass substrate (step C, Figure 2). Next (step D, Figure 2), the sample is immersed in acetone for 3 hours to dissolve the MMA layer and release the Si wafer, resulting in structures where Au nanorods are partially buried within $TiO_2$ (step E, Figure 2). For the experiments to be described below, we have also prepared a reference sample consisting of *non-embedded* Au nanorods on $TiO_2$. In this case, the self–assembly of Au nanorods is performed on the surface of $TiO_2$ after the acetone lift–off step (step E, Figure 2).

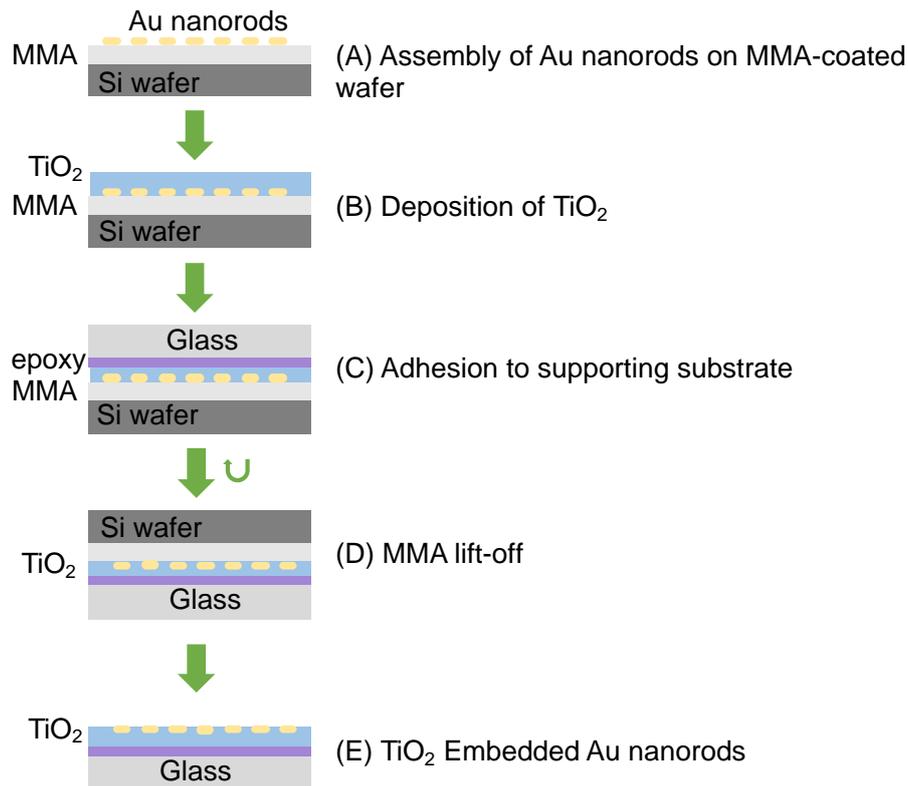

**Figure 2.** *Nanofabrication. (A) Methyl methacrylate (MMA) is spin–coated on a Si wafer and Au nanorods are self–assembled on the MMA layer. (B) Deposition of 50 nm TiO$_2$ via electron beam evaporation. (C) The sample is adhered to a glass substrate via an epoxy layer and subsequently in step (D) it is immersed (upside-down) in acetone to remove the MMA layer and release the structure from Si wafer to form the final (E) partially embedded structure.*

Figure 3(A) and (B) show surface topography images obtained with Atomic Force Microscopy (AFM) for the resulting embedded and non–embedded structures. In Figure 3(A), the Au nanorods are embedded within the TiO$_2$ and the AFM topography image consists of a random collection of pits denoting the location where the Au nanorods are embedded. In fact, the nanorods are shown to be buried ~ 2 nm below the TiO$_2$ surface (supporting information figure S2). On the contrary, the surface topography of Figure 3(B) clearly reveals the geometrical outline of the Au nanorods that rest on top of the surface of the TiO$_2$ film. Figures 3(C) and (D) show scanning electron microscopy images of both kinds of structures [3(C):embedded, and 3(D): non–embedded] where the presence of Au nanorods on both samples is clearly visible. Together, the images of Figure 3 indicate that the process described in Figure 2 leads to structures where Au nanorods are partially buried inside TiO$_2$.

Figure 3(E) shows the measured optical extinction [defined as $-\ln(T)$, $T$ is the transmission spectrum] for the structures of Figure 3, along with the measured absorption spectrum for the Au nanorods in solution. The spectrum in solution is dominated by a single band that corresponds to the longitudinal localised surface plasmon resonance of the Au nanorods. The extinction spectra of the Au nanorods when placed on top and when partially buried inside a thin film of TiO$_2$ show redshifts, which are attributed to the presence of the high refractive index of the TiO$_2$. Numerical full wave simulations were carried out to identify the observed extinction bands, as shown in Figure 3(F). In figure 3(F) we show the calculated absorption spectra for idealised structures consisting of spherically–capped Au nanorods resting on top of a thin film with refractive index 2.4, and for a similar structure embedded to a variable extent in the film. By immersing the Au nanorod in the semiconductor, the localised surface plasmon resonances (both longitudinal and transverse) redshift in proportion to the amount of surface coverage.

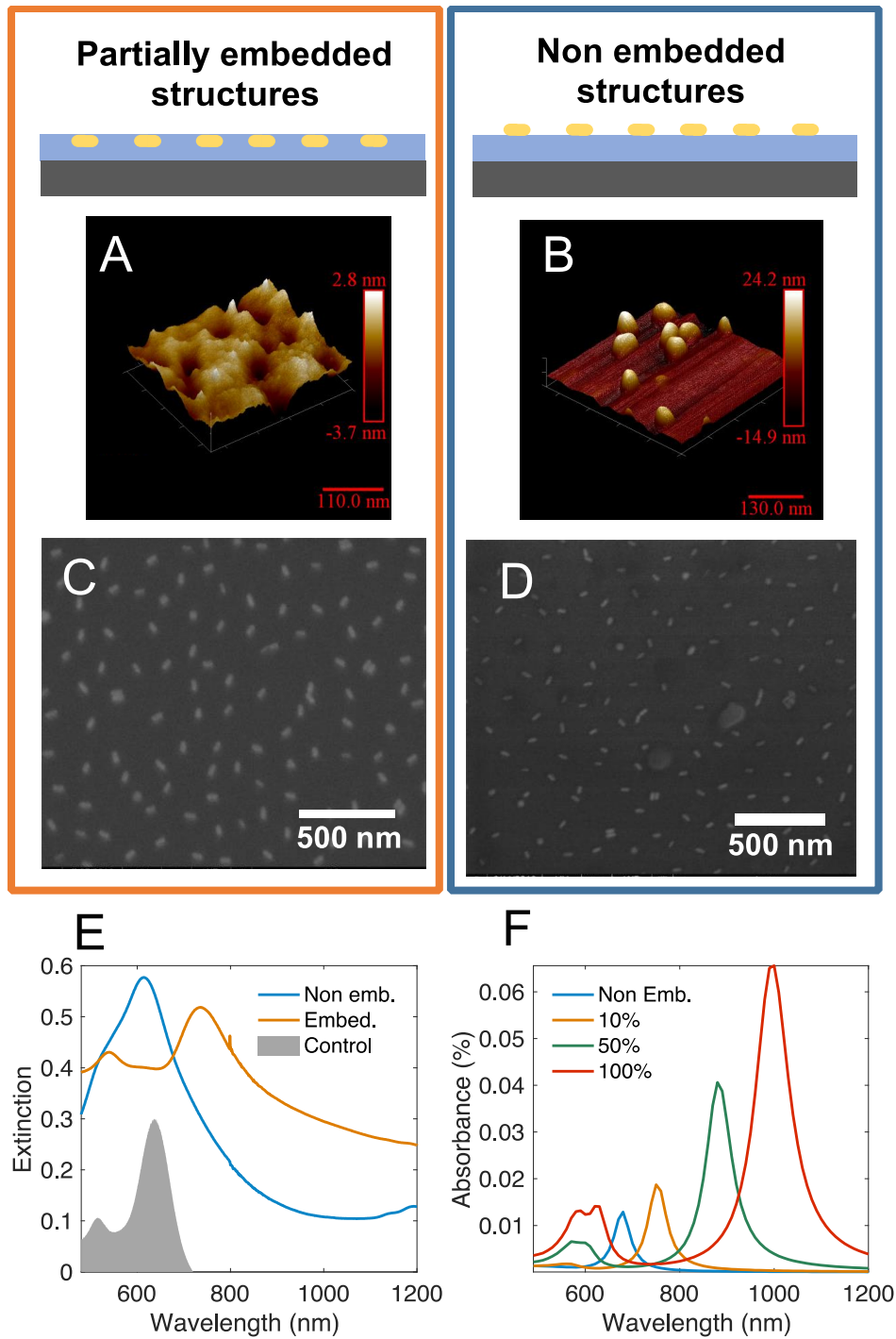

**Figure 3.** *Structural characterisation and Optical properties. (A) and (B) are atomic force microscopy topography images of the embedded and non-embedded structures respectively. (C) and (D) are the corresponding scanning electron images. (E) Measured extinction [−ln(T)] spectra of the embedded, non-embedded structures along with the absorption spectrum of the Au nanorods in solution. (F) Calculated absorption spectra using a full-wave numerical solution of Maxwell equations.*

Following Landau (non–radiative) damping of surface plasmon resonances, the resulting hot–electrons have a finite probability of injection into the semiconductor material[10]. Once injected, these carriers rapidly relax to the lower states of the conduction band of the semiconductor, and can thus absorb both visible and near–infrared light enabling the detection of these charge–transfer processes through visible pump near–IR probe spectroscopy, as we describe next.[21,22]

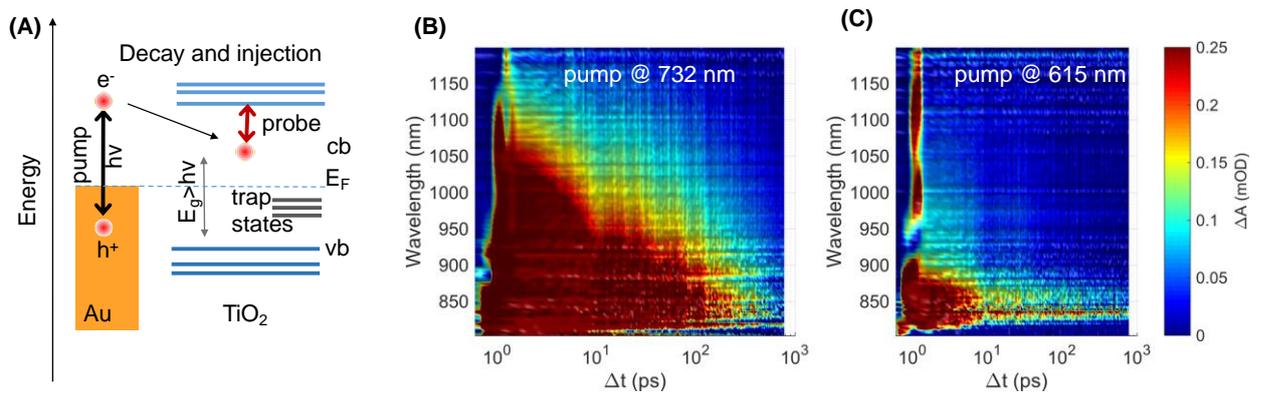

**Figure 4.** *Ultra--fast pump probe spectroscopy. (A) Pump pulses excite localised surface plasmon resonances in the Au nanorods which, after non--radiative decay, can result in the injection of hot--electrons into the conduction band (cb) of $TiO_2$. Intraband excitation of these electrons is possible with NIR probe pulses. Also shown are the relevant energy levels of the system: $E_F$: Fermi level of the metal, vb: valence band and $E_g$: semiconductor bandgap. (B) and (C) show the transient absorption ΔA spectra as a function of probe wavelength and pump--probe delay time (Δt) for the (B) partially and the (C) non--embedded structures. The pump wavelengths are indicated in the figures.*

As shown in the diagram of Figure 4(A), we photo–excited the samples using ultra–short light pulses spectrally tuned to the localised surface plasmon resonance of the Au nanorods, which according to the data of Figure 3(E), was 634 nm for the non–embedded structures and 732 nm for the embedded structures. These wavelengths are significantly below the band-gap of $TiO_2$ and consequently, no direct excitation of electron-hole pairs in this material is possible. After non–radiative relaxation of the localised surface plasmon resonances, injection of electrons into the TiO$_2$ can take place, leading to a transient population of electrons at the bottom of the conduction band of the semiconductor [Figure 4(A)]. These electrons can absorb near–infrared light (via intraband electronic transitions) and are thus detectable through the changes in the absorption of time–delayed near–IR light probe pulses.[22]

Figure 4(B) and (C) show the results of these measurements as a function of probe wavelength and pump–probe time delay. In general, the measured transient absorption ($\Delta A$) spectra indicate excited state absorption ($\Delta A > 0$), where the amplitudes of the measured signals are larger for the samples where the Au nanorods are partially embedded in TiO$_2$. In the time domain, the detected signals consist of a fast rise (sub–picosecond time–scale) and a rapid decay ($\Delta t < 5$ ps) which is followed by a slower, multi–exponential decay ($\Delta t > 5$ ps). It has been documented that the electron population in TiO$_2$ decreases with time due to an interplay of processes such as: recombination with the positive holes present in Au[23], surface electron trapping ($\Delta t < 200$ fs, Figure 5 shows a representation of such states)[24] and charge transfer to ambient oxygen molecules ($10^{-6} - 10^{-3}$s)[25], processes with marked differences in time scales, which can account for the multi–rate decay. The transient absorption changes measured for partially embedded Au nanorods in metal–semiconductor–nanorod structures (Figure S4), exhibit damped oscillations with a strong Fourier component at ∼ 20 GHz (supporting information sections S2 and S3). We assign this oscillation frequency to mechanical (breathing mode) oscillations of the nanorods. These oscillations are part of the non–radiative decay channels available to localised surface plasmon resonances in the Au nanorods[14], and occur in competition with electron injection across the Au–TiO$_2$ interface.

In Figure 5(A), we show a direct comparison of the transient absorption changes for Au nanorods that are partially embedded, non-embedded and a control sample where the nanorods are deposited directly on a glass substrate. For the rods on a glass substrate, no transient signal is observed, which indicates that no electron injection events take place, a result consistent with the expectation that few or no electron-accepting states exist in glass. On the contrary, for the other two cases there is a non-negligible transient signal, which is largest (in magnitude) for the partially embedded samples. We argue that this difference in signal magnitude is largely due to a higher probability of electron injection afforded by the larger number-density of Au–TiO$_2$ interfaces in the partially embedded structures.

We now estimate the changes in hot-electron injection quantum yield using the data derived from the transient absorption spectroscopy. In the small signal limit, the magnitude of the transient absorption signal (that originates from the electrons residing in the semiconductor) is proportional to the number of photons absorbed by the metal nanostructure.[8,15] The proportionality constant is given by a product of the absorption cross-section and the photon-to-injected-electron quantum yield (supporting information section S4). As shown in the supporting information section (Figure S8), the magnitude of the measured transient absorption increases linearly with pump power, indicating that the experiments were carried out in the small signal limit. We can therefore quantify the increase in hot-electron injection quantum yield $R_{QY}$, by comparing the magnitude of the measured transient absorption signals for the embedded and non–embedded samples (taking into account the existing differences in the amount of absorbed pump photons)[8,15]:

$$R_{QY} = \frac{\left(\frac{\Delta A_{emb,732nm}(1000nm)}{(1-10^{-OD})j_{(732nm)}}\right)}{\left(\frac{\Delta A_{non-emb,615nm}(1000nm)}{(1-10^{-OD})j_{(615nm)}}\right)},$$

where $\Delta A_{x,pump}(probe)$ is the measured peak amplitude of the transient change in absorbance at the *probe* wavelength, for sample *x*, irradiated with the *pump* wavelength (these values were averaged between 0.2 and 0.4 ps). $j$ is the incident photon flux and $OD$ the measured optical density for absorption of the pump pulses. Using this equation with the data of Figure 4, we obtain a value of $R_{QY}$ = 2.7 indicating that the method outlined in Figure 2 leads to structures with increased hot–electron–injection.

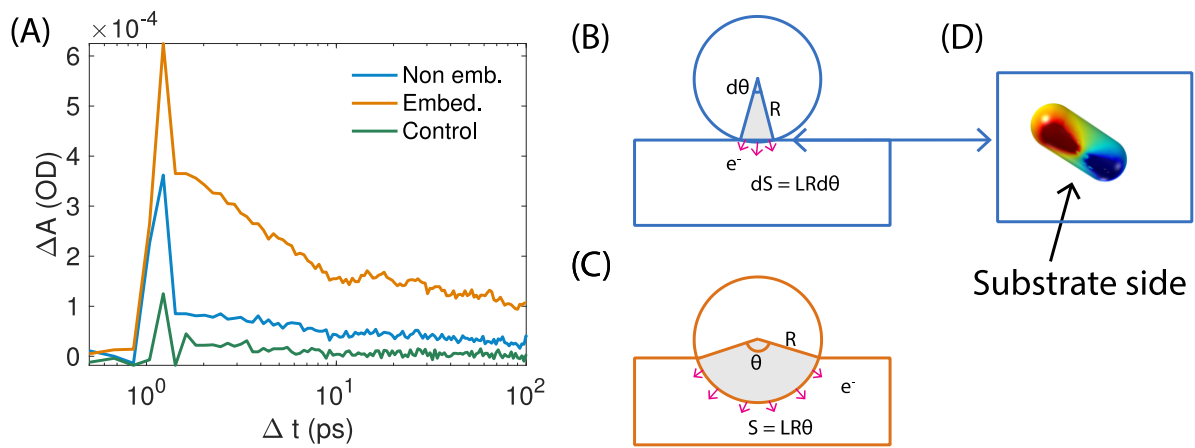

**Figure 5.** *Amplitude of transient absorption signal. (A) Time evolution of the transient absorption ΔA signal measured at 1000 nm for samples consisting of Au nanorods on glass, on TiO₂ and partially–embedded in TiO₂. Also shown are diagrams of the idealised cross-section of cylindrical Au nanorods of radius R and length L (B) on TiO₂ and (C) partially embedded in it. The arrows indicate interface normal vectors. In (B), the Au–TiO₂ contact area is dS, whereas this area increases to S in the case shown in (C). (D) Map of*

*the electric field for a nanorod placed on top of a TiO$_2$ substrate. The colours indicate the relative sign of the field.*

Having estimated the increase in injection quantum yield, we now discuss if the observed value of $R_{QY}$ is consistent with an increase in the metal–semiconductor contact area resulting from partially embedding the nanorods. $R_{QY}$ can increase with contact area if the population of hot–electrons resulting from plasmon relaxation possesses an isotropic energy–momentum distribution, for in this case the probability for electron injection is partly determined by the flux of these charge carriers through the metal–semiconductor contact.[26,27] In Figure 5(B)-(C), we show idealised cross–sections of a cylindrical nanorod (of radius R and length L) placed either on (B) top or (C) embedded in a semiconductor material. Within this idealised geometry, the ratio of contact areas $R_A$ is given by $R_A = S/dS = \theta/d\theta$, where $\theta$ is an angle describing the fraction of the cylinder making contact with the semiconductor. Given that $d\theta \ll \theta$, $R_A$ can take values that can significantly exceed the estimated ratio of injection quantum yields $R_{QY}$. Possible reasons for the discrepancy include: (1) the organic molecules that decorate the Au nanorods hinder the charge transfer events[28], (2) poor crystallinity of the TiO$_2$, leading to trap states[29] [Figure 4(A)] which quickly trap the injected carriers, (3) non–conformal coating of the surface of the Au nanorods with TiO$_2$ during physical vapour deposition, or (4) the assumption of spatial invariance of the energy–momentum distribution of hot–electrons is not valid, and this distribution is instead non–uniform.[9,30,31] Points (1)-(3) highlight some of the limitations of our nanofabrication approach. In particular, it is known that an intense annealing process is required to improve the material properties of the semiconductor[32], but this process will likely melt the Au nanorods. A possible explanation for point (4) is that electron injection into TiO$_2$ occurs via a *surface mechanism*:[33] a directional emission process wherein an electron moving towards the metal–semiconductor interface absorbs the energy of a photon and thus overcomes the potential energy barrier that exists at the interface.

In the surface charge emission mechanism that occurs via a plasmon excitation, the rate of electron injection into the semiconductor is proportional to the square of the absolute value of the electric field normal to the metal–semiconductor interface (supplementary information section S6). Spatial maps of these fields, shown in Figure 5(D), clearly demonstrate their spatially inhomogeneous nature, and in particular, the strong confinement of electromagnetic fields that exists at the metal–semiconductor interface. This strong localisation, according to the mentioned mechanism, implies that increases in the surface coverage of the nanostructure would lead to enhancements in the quantum yield of hot–electron injection that are smaller than the corresponding increase in surface contact area. In fact, our calculations show that for the geometry of our nanostructures, the largest enhancement is expected to be in the order of 3.26×, which is of a similar order of magnitude to our experimental results. This puts forward evidence indicating that the non–radiative plasmon decay and subsequent electron injection into the semiconductor does not occur via the formation of an intermediate hot electron gas with an isotropic momentum distribution, but instead, the energy of the coherent oscillation of charge density (localised surface plasmon resonance) can be transferred to a charge carrier moving to the interface; a directional process.

It is important to note that an estimate of the injection quantum yield for each structure could be made by comparing the measured $\varDelta A$ signals to those obtained by directly populating the TiO$_2$ conduction band via transient absorption measurements using pump pulses with energies above the bandgap of the semiconductor[34] ($E_g \approx 3.2$ eV for TiO$_2$[35]).[15] Our instrument is currently not capable of producing these pump pulses.

In summary, we have shown that the increase in metal–semiconductor contact area leads to a higher efficiency of hot–electron injection into the semiconductor. Detailed analysis of our data suggests that surface charge emission mechanism that occurs via a plasmon excitation is likely to be the dominant mechanism for hot–electron injection. In order to increase the metal–semiconductor contact area, we have developed a fabrication process that only requires physical vapour deposition techniques and does not involve any complex processes such as electron beam lithography or reactive ion etching, making it highly attractive for large scale fabrication. In particular, we envisage that contrary to other approaches[36,37], our method can enable the construction of near–perfect absorbers of light that incorporate partially-buried metal nanostructures.[18] We also envisage that these substrates can be used as active elements in novel sensors[38], photo–detectors[39] or in plasmonic photochemistry.[5,35,40–45]

**Acknowledgements**

This work was performed in part at the Melbourne Centre for Nanofabrication (MCN) in the Victorian Node of the Australian National Fabrication Facility (ANFF). C. N. was supported by an OCE Fellowship from CSIRO. D.E.G. acknowledges the ARC for support through a Future Fellowship **(FT140100514)**. D.E.G and U.B. acknowledge the ANFF for the MCN Technology Fellowships. The authors acknowledge use of facilities within the Monash Centre for Electron Microscopy A.R., T.J.D. & D.E.G. acknowledge the ARC for support through a Discovery Project (**DP160100983**). T.A.S., D.E.G. U.B. and J.E.S. also acknowledge support from the Australian Research Council Centre of Excellence in Exciton Science (**CE170100026**).

**Supporting Information Available:** Detailed experimental section, further material characterization, mirror-semiconductor-nanorod structures, Assignment of mechanical oscillations and Kinetic model

**References**

(1) Sze, S., Ng, K. *Physics of semiconductor devices*; Wiley-interscience Publication; John Wiley & Sons, 1981.

(2) Zhang, Z.; Yates, J. T. Band bending in semiconductors: Chemical and physical consequences at surfaces and interfaces. *Chem. Rev.* **2012**, *112* (10), 5520–5551.

(3) Knight, M. W.; Sobhani, H.; Nordlander, P.; Halas, N. J. Photodetection with active optical antennas. *Science* **2011**, *332* (6030), 702–704.

(4) Reineck, P.; Lee, G. P.; Brick, D.; Karg, M.; Mulvaney, P.; Bach, U. A solid-state plasmonic solar cell via metal nanoparticle self-assembly. *Adv. Mater.* **2012**, *24* (35), 4750–4755.

(5) Xiao, Q.; Sarina, S.; Bo, A.; Jia, J.; Liu, H.; Arnold, D. P.; Huang, Y.; Wu, H.; Zhu, H. Visible light-driven cross-coupling reactions at lower temperatures using a photocatalyst of palladium and gold alloy nanoparticles. *ACS Catal.* **2014**, *4* (6), 1725–1734.

(6) Takahashi, Y.; Tatsuma, T. Solid state photovoltaic cells based on localized surface plasmon-induced charge separation. *Appl. Phys. Lett.* **2011**, *99,* 182110.

(7) Mubeen, S.; Lee, J.; Singh, N.; Kramer, S.; Stucky, G. D.; Moskovits, M. An autonomous photosynthetic device in which all charge carriers derive from surface plasmons. *Nat. Nanotechnol.* **2013**, *8* (4), 247–251.

(8) Wu, K.; Rodríguez-Córdoba, W. E.; Yang, Y.; Lian, T. Plasmon-induced hot electron transfer from the au tip to CdS rod in CdS-au nanoheterostructures. *Nano Lett.* **2013**, *13* (11), 5255-5263.

(9) Knight, M. W.; Wang, Y.; Urban, A. S.; Sobhani, A.; Zheng, B. Y.; Nordlander, P.; Halas, N. J. Embedding plasmonic nanostructure diodes enhances hot electron emission. *Nano Lett.* **2013**, *13*, 1687.

(10) White, T. P.; Catchpole, K. R. Plasmon-enhanced internal photoemission for photovoltaics: Theoretical efficiency limits. *Appl. Phys. Lett.* **2012**, *101* (7), 073905.

(11) Brongersma, M. L.; Halas, N. J.; Nordlander, P. Plasmon-induced hot carrier science and technology. *Nat. Nanotechnol.* **2015**, *10* (1), 25–34.

(12) Giugni, A.; Torre, B.; Toma, A.; Francardi, M.; Malerba, M.; Alabastri, A.; Proietti Zaccaria, R.; Stockman, M. I.; Di Fabrizio, E. Hot-electron nanoscopy using adiabatic compression of surface plasmons. *Nat. Nanotechnol.* **2013**, *8*, 845-852.

(13) Scales, C.; Berini, P. Thin-film schottky barrier photodetector models. *Quantum Electronics, IEEE Journal of* **2010**, *46* (5), 633–643.

(14) Hartland, G. V. Optical studies of dynamics in noble metal nanostructures. *Chem. Rev.* **2011**, *111* (6), 3858–3887.

(15) Wu, K.; Chen, J.; McBride, J. R.; Lian, T. Efficient hot-electron transfer by a plasmon-induced interfacial charge-transfer transition. *Science* **2015**, *349* (6248), 632–635.

(16) Shaviv, E.; Schubert, O.; Alves-Santos, M.; Goldoni, G.; Di Felice, R.; Vallée, F.; Del Fatti, N.; Banin, U.; Sönnichsen, C. Absorption properties of metal–Semiconductor hybrid nanoparticles. *ACS Nano* **2011**, *5* (6), 4712–4719.

(17) Kim, M.; Lin, M.; Son, J.; Xu, H.; Nam, J.-M. Hot-electron-mediated photochemical reactions: Principles, recent advances, and challenges. *Adv. Opt. Mater.* **2017,** *5*(15), 1700004.

(18) Ng, C.; Cadusch, J.; Dligatch, S.; Roberts, A.; Davis, T. J.; Mulvaney, P.; Gomez, D. E. Hot carrier extraction with plasmonic broadband absorbers. *ACS Nano* **2016**, *10*, 4704—4711.

(19) Robatjazi, H.; Bahauddin, S. M.; Doiron, C.; Thomann, I. Direct plasmon-driven photoelectrocatalysis. *Nano Lett.* **2015,** *15*(9), 6155-6161.

(20) Fang, Y.; Jiao, Y.; Xiong, K.; Ogier, R.; Yang, Z.; Gao, S.; Dahlin, A.; Käll, M. Plasmon enhanced internal photoemission in antenna-spacer-mirror based $Au/TiO_2$ nanostructures. *Nano Lett.* **2015,** *15*(6), 4059-4065.

(21) Tachikawa, T.; Tojo, S.; Fujitsuka, M.; Sekino, T.; Majima, T. Photoinduced charge separation in titania nanotubes. *J. Phys. Chem. B* **2006**, *110* (29), 14055–14059.

(22) Zeng, P.; Cadusch, J.; Chakraborty, D.; Smith, T. A.; Roberts, A.; Sader, J.; Davis, T.; Gomez, D. E. Photo-induced electron transfer in the strong coupling regime: Waveguide-plasmon polaritons. *Nano Lett.***2016**, *16*, 2651–2656.

(23) Bian, Z.; Tachikawa, T.; Zhang, P.; Fujitsuka, M.; Majima, T. $Au/TiO_2$ superstructure-based plasmonic photocatalysts exhibiting efficient charge separation and unprecedented activity. *J. Am. Chem. Soc.* **2014**, *136* (1), 458–465.

(24) Tamaki, Y.; Furube, A.; Murai, M.; Hara, K.; Katoh, R.; Tachiya, M. Dynamics of efficient electron-hole separation in $TiO_2$ nanoparticles revealed by femtosecond transient absorption spectroscopy under the weak-excitation condition. *Phys. Chem. Chem. Phys.* **2007**, *9* (12), 1453–1460.

(25) Yoshihara, T.; Katoh, R.; Furube, A.; Tamaki, Y.; Murai, M.; Hara, K.; Murata, S.; Arakawa, H.; Tachiya, M. Identification of reactive species in photoexcited nanocrystalline TiO$_2$ films by wide-wavelength-range (400–2500 nm) transient absorption spectroscopy. *J. Phys. Chem. B* **2004**, *108* (12), 3817–3823.

(26) Chen, Q. Y.; Bates, C. W. Geometrical factors in enhanced photoyield from small metal particles. *Phys. Rev. Lett.* **1986**, *57* (21), 2737–2740.

(27) Müller, U.; Burtscher, H.; Schmidt-Ott, A. Photoemission from small metal spheres: A model calculation using an enhanced three-step model. *Phys. Rev. B* **1988**, *38* (11), 7814–7816.

(28) Smith, J. G.; Jain, P. K. The ligand shell as an energy barrier in surface reactions on transition metal nanoparticles. *J. Am. Chem. Soc.* **2016**, *138* (21), 6765–6773.

(29) Kodiyath, R.; Manikandan, M.; Liu, L.; Ramesh, G. V.; Koyasu, S.; Miyauchi, M.; Sakuma, Y.; Tanabe, T.; Gunji, T.; Duy Dao, T.; et al. Visible-light photodecomposition of acetaldehyde by TiO$_2$-coated gold nanocages: Plasmon-mediated hot electron transport via defect states. *Chem. Comm.* **2014**, *50* (98), 15553–15556.

(30) Shalaev, V. M.; Douketis, C.; Stuckless, J. T.; Moskovits, M. Light-induced kinetic effects in solids. *Phys. Rev. B* **1996**, *53* (17), 11388–11402.

(31) Petek, H. Photoexcitation of adsorbates on metal surfaces: One-step or three-step. *J. of Chem. Phys.* **2012**, *137* (9), 091704.

(32) Zheng, B. Y.; Zhao, H.; Manjavacas, A.; McClain, M.; Nordlander, P.; Halas, N. J. Distinguishing between plasmon-induced and photoexcited carriers in a device geometry. *Nat. Commun.* **2015**, *6, 7797*.

(33) Babicheva, V. E.; Zhukovsky, S. V.; Ikhsanov, R. S.; Protsenko, I. E.; Smetanin, I. V.; Uskov, A. Hot electron photoemission from plasmonic nanostructures: The role of surface photoemission and transition absorption. *ACS Photonics* **2015,** *2*(8), 1039-1048.

(34) Ratchford, D. C.; Dunkelberger, A. D.; Vurgaftman, I.; Owrutsky, J. C.; Pehrsson, P. E. Quantification of efficient plasmonic hot-electron injection in gold nanoparticle–TiO$_2$ films. *Nano Lett.* **2017,** *17*(10), 6047-6055.

(35) Zhang, X.; Chen, Y. L.; Liu, R.-S.; Tsai, D. P. Plasmonic photocatalysis. *Rep. Prog. Phys.* **2013**, *76* (4), 046401.

(36) Wu, B.; Liu, D.; Mubeen, S.; Chuong, T. T.; Moskovits, M.; Stucky, G. D. Anisotropic growth of TiO$_2$ onto gold nanorods for plasmon-enhanced hydrogen production from water reduction. *J. Am. Chem. Soc.* **2016**, *138* (4), 1114–1117.

(37) Mubeen, S.; Lee, J.; Liu, D.; Stucky, G. D.; Moskovits, M. Panchromatic photoproduction of H$_2$ with surface plasmons. *Nano Lett.* **2015**, *15*, 2132-2136.

(38) Chen, X.; Ciracì, C.; Smith, D. R.; Oh, S.-H. Nanogap-enhanced infrared spectroscopy with template-stripped wafer-scale arrays of buried plasmonic cavities. *Nano Lett.* **2015**, *15* (1), 107–113.

(39) Li, W.; Valentine, J. Metamaterial perfect absorber based hot electron photodetection. *Nano Lett.* **2014**, *14* (6), 3510–3514.

(40) Mukherjee, S.; Zhou, L.; Goodman, A. M.; Large, N.; Ayala-Orozco, C.; Zhang, Y.; Nordlander, P.; Halas, N. J. Hot-electron-induced dissociation of $H_2$ on gold nanoparticles supported on $SiO_2$. *J. Am. Chem. Soc.* **2014**, *136* (1), 64–67.

(41) Ke, X.; Zhang, X.; Zhao, J.; Sarina, S.; Barry, J.; Zhu, H. Selective reductions using visible light photocatalysts of supported gold nanoparticles. *Green Chem.* **2013**, *15* (1), 236–244.

(42) Marimuthu, A.; Zhang, J.; Linic, S. Tuning selectivity in propylene epoxidation by plasmon mediated photo-switching of Cu oxidation state. *Science* **2013**, *339* (6127), 1590–1593.

(43) Sarina, S.; Zhu, H.; Jaatinen, E.; Xiao, Q.; Liu, H.; Jia, J.; Chen, C.; Zhao, J. Enhancing catalytic performance of palladium in gold and palladium alloy nanoparticles for organic synthesis reactions through visible light irradiation at ambient temperatures. *J. Am. Chem. Soc.* **2013**, *135* (15), 5793–5801.

(44) Scaiano, J. C.; Stamplecoskie, K. Can surface plasmon fields provide a new way to photosensitize organic photoreactions? From designer nanoparticles to custom applications. *J. Phys. Chem. Lett.* **2013**, *4* (7), 1177–1187.

(45) Christopher, P.; Xin, H.; Linic, S. Visible-light-enhanced catalytic oxidation reactions on plasmonic silver nanostructures. *Nat. Chem.* **2011**, *3* (6), 467–472.